\documentclass[journal]{IEEEtran}

\usepackage{cite}
\usepackage{bm}

\ifCLASSINFOpdf
  \usepackage[pdftex]{graphicx}
  \graphicspath{{./Images/}}
  \DeclareGraphicsExtensions{.pdf,.jpeg,.png}
\else
\fi

\usepackage[cmex10]{amsmath}
\allowdisplaybreaks
\usepackage{url}
\usepackage[draft]{hyperref}
\usepackage{amssymb}
\usepackage{multirow}
\usepackage{booktabs,colortbl}
\usepackage{xcolor}
\usepackage{siunitx}
\usepackage{psfrag}
\usepackage{commath}

\usepackage{cite}
\usepackage{amsmath,amssymb,amsfonts}
\usepackage{algorithmic}
\usepackage{graphicx}
\usepackage{textcomp}
\usepackage{flushend} 

\usepackage[nonumberlist,xindy]{glossaries}
\usepackage{enumitem}
\usepackage{mathtools}
\def\BibTeX{{\rm B\kern-.05em{\sc i\kern-.025em b}\kern-.08em
    T\kern-.1667em\lower.7ex\hbox{E}\kern-.125emX}}

\newacronym{opf}{OPF}{optimal power flow}
\newacronym{res}{RES}{renewable energy sources}
\newacronym{eds}{EDS}{electrical distribution systems}
\newacronym{milp}{MILP}{mixed-integer linear programming}
\newacronym{minlp}{MINLP}{mixed-integer nonlinear programming}
\newacronym{misocp}{MISOCP}{mixed-integer second-order cone programming}
\newacronym{dg}{DG}{distributed generator}
\newacronym{der}{DER}{distributed energy resource}
\newacronym{ess}{ESS}{Energy storage systems}
\newacronym{mcs}{MCS}{Monte Carlo simulation}
\newacronym{wt}{WT}{wind turbine}
\newacronym{pv}{PV}{Photovoltaic}
\newacronym{soc}{SOC}{state of charge}
\newacronym{pdf}{PDF}{probability density function}
\newacronym{PDF}{PDF}{Probability density functions}
\newacronym{cdf}{CDF}{cumulative distribution function}
\newacronym{oem}{OEM}{optimal energy management}
\newacronym{vsc}{VSC}{voltage-source converter}
\newacronym{csc}{CSC}{current-source converter}
\newacronym{vpl}{VPL}{virtual power line}

\setglossarystyle{super}

\definecolor{yaxis}{rgb}{0.85, 0.33, 0.10}
\definecolor{new_gray}{rgb}{0.026, 0.024, 0.053}

\usepackage{geometry}   
\begin{document}

\title{{Estimating Risk-Aware} Flexibility Areas for EV Charging Pools via Stochastic AC-OPF} 

\author{\IEEEauthorblockN{Juan S. Giraldo, \IEEEmembership{Member, IEEE}, 
Nataly Ba\~{n}ol Arias, \IEEEmembership{Member, IEEE}, 
Pedro P. Vergara, \IEEEmembership{Member, IEEE}, \\ 
Maria Vlasiou, 
Gerwin Hoogsteen, \IEEEmembership{Member, IEEE}, 
and Johann L. Hurink, \IEEEmembership{Member, IEEE}
}
\thanks{
This work was financially supported by the Netherlands Enterprise Agency (RVO) -- DEI+ project~{120037} ``\textit{Het Indië terrein: Een slimme buurtbatterij in de oude weverij}". }
\thanks{J. S. Giraldo is with the Energy Transition Studies group, Netherlands Organisation for Applied Scientific Research (TNO), Amsterdam, 1043~NT, The Netherlands (e-mail: juan.giraldo@tno.nl).}
\thanks{N. B. Arias, M. Vlasiou, G. Hoogsteen, and J. L. Hurink are with the dept. Electrical Engineering, Mathematics and Computer Science, University of Twente, Enschede, 7522 NB, The Netherlands, (e-mail:\{m.n.banolarias; m.vlasiou; g.hoogsteen; j.l.hurink\}@utwente.nl).}
\thanks{P. P. Vergara is with the Intelligent Electrical Power Grids group, EEMCS, Delft University of Technology, Delft, 2628 CD, The Netherlands, (e-mail: p.p.vergarabarrios@tudelft.nl).}
}

\maketitle

\begin{abstract}

This paper introduces a stochastic AC-OPF (SOPF) for the flexibility management of electric vehicle (EV) charging pools in distribution networks under uncertainty. The SOPF considers discrete utility functions from charging pools as a compensation mechanism for eventual energy not served to their charging tasks. An application of the proposed SOPF is described where a distribution system operator (DSO) requires flexibility to each charging pool in a day-ahead time frame, minimizing the cost for flexibility while guaranteeing technical limits. 
Flexibility areas are defined for each charging pool and calculated as a function of a risk parameter involving the solution's uncertainty. 
Results show that all players can benefit from this approach, i.e., the DSO obtains a risk-aware solution, while charging pools/tasks perceive a reduction in the total energy payment due to flexibility services.     

\end{abstract}


%


\section*{Nomenclature}
\noindent \emph{Sets}
\begin{description}[itemsep=0.2ex, labelwidth=1.2cm, leftmargin=1.4cm]
\item[$\Omega_{b}$] Set of nodes
\item[$\Omega_{\mathrm{S}}$] Set of nodes with charging pools
\item[$\Omega_{\mathrm{N}}^{s}$] Set of charging points at charging pool $s\in\Omega_{\mathrm{S}}$
\item[$\Omega_{\mathrm{K}}^{s}$] Set of breaking points at charging pool $s\in\Omega_{\mathrm{S}}$
\item[$\Omega_{\mathrm{T}}$] Set of time periods
\item[$\Omega_{\omega}$] Set of stochastic scenarios
\end{description}
\noindent \emph{Parameters}
\begin{description}[itemsep=0.2ex, labelwidth=1.2cm, leftmargin=1.4cm]
\item[$\mathrm{A}_{n,\omega}$] Characteristics of charging task $n$ at scenario $\omega$
\item[$\mathrm{a}_{n,\omega}$] Arrival time of charging task $n$ at scenario $\omega$
\item[$\mathrm{d}_{n,\omega}$] Departure time of charging task $n$ at scenario $\omega$
\item[$\mathrm{E}_{n,\omega}$] Required energy of charging task $n$ at scenario $\omega$
\item[$\mathrm{R}_{ij},\,\mathrm{X}_{ij}$] Resistance and reactance of branch connecting nodes $ij$
\item[$\mathrm{P}^{\rm{D}}_{i,t},\,\mathrm{Q}^{\rm{D}}_{i,t}$] Active and reactive demand power at node $i$ and period $t$
\item[$\eta_{n}^{\mathrm{a}}$] Expected arrival time of charging task $n$
\item[$\eta_{n}^{\mathrm{d}}$] Expected departure time of charging task $n$
\item[$\beta_{s,t}$] Risk parameter at charging pool $s$ in period $t$
\item[$\kappa$] Number of breaking points
\item[$\overline{\mathrm{p}}_{s,t}$] Maximum power allowed of charging pool $s$ in period $t$
\item[$\overline{\mathrm{x}}_{n}$] Maximum charging power of charging task $n$
\item[$\mathrm{h}_{s,k},\,\mathrm{b}_{s,k}$] Coefficients of the utility function at charging pool $s$ and break point $k$
\item[$\alpha_{s,k}$] Break point value of energy not served at pool $s$ and point $k$
\item[$\Delta\mathrm{t}$] Duration of the time period $t$
\item[$\mathrm{c}_{s,t}$] Unitary cost of energy at charging pool $s$ period $t$
\item[$\overline{\rm{I}}_{ij}$] Maximum allowed current magnitude at branch $i$-$j$
\item[$\overline{\rm{V}}$] Maximum allowed voltage magnitude 
\item[$\underline{\rm{V}}$] Minimum allowed voltage magnitude 
\item[$\pi_{\omega}$] Probability of scenario $\omega$
\end{description}

\noindent \emph{Variables}
\begin{description}[itemsep=0.2ex, labelwidth=1.4cm, leftmargin=1.6cm]
\item[$p_{s,t}$] Reserved power for charging pool $s$ in period $t$
\item[$x_{s,t,\omega}$] Allocated power consumption for charging pool $s$ in period $t$ and scenario $\omega$
\item[$\rho_{s,t,\omega}$] Power mismatch for charging pool $s$ in period $t$ and scenario $\omega$
\item[$\phi_{n,\omega}$] Energy not served to task $n$ in scenario $\omega$
\item[$\Phi_{s,\omega}$] Total energy not served at charging pool $s$ in scenario $\omega$
\item[$\underline{\lambda}_{s,k,\omega},\,\overline{\lambda}_{s,k,\omega}$] Weights in break point $k$, at pool $s$ in scenario $\omega$ 
\item[$P_{ij,t,\omega},\,Q_{ij,t,\omega}$] Active and reactive power flowing through branch $i$-$j$ in period $t$ in scenario $\omega$ 
\item[$I^{\rm{sqr}}_{ij,t,\omega}$] Squared current magnitude flowing through branch $i$-$j$ in period $t$ in scenario $\omega$ 
\item[$V^{\rm{sqr}}_{i,t,\omega}$] Squared voltage magnitude at node $i$ in period $t$ in scenario $\omega$
\item[$y_{s,k,\omega}$] Binary variable representing state of segment at break point $k$, pool $s$ in scenario $\omega$
\item[$\mathcal{R}_{s,t}$] Flexibility area of charging pool $s$ in period $t$
\item[$\mathcal{Z}_{s,\omega}$] Cost for energy not served at charging pool $s$ and scenario $\omega$

\end{description}

\section{Introduction}


\IEEEPARstart{B}{esides} being an environmentally-friendly option for transportation, electric vehicles (EVs) can also provide services due to the controllable nature of their load. Examples of these services are, amongst others, congestion management, peak shaving, and frequency regulation~\cite{arias2019distribution}. These services may be of increased value as technical problems, such as voltage violations and branch overloading, are expected to be more likely in distribution systems if no actions are taken~\cite{hoogsteen2017charging}. For distribution system operators (DSOs), which are responsible for delivering electricity to end customers and maintaining a reliable network operation, it might be interesting to assess the flexibility needs in their networks. In a later stage, these flexibility needs may also be provided by entities such as aggregators to solve operational issues or offer it as an ancillary service. For this, flexibility areas may be determined corresponding to the range of active power in which flexibility sources can be managed~\cite{silva2018estimating}. 

 
In~\cite{limmer2019peak}, it is already stated that network issues can be tackled through flexibility management frameworks to avoid common issues in distribution systems, such as congestion or voltage limit violations. This strategy is known as {DSO's flexibility procurement}, and it has gained momentum during the last few years due to its economic advantages over other solutions such as grid reinforcement. However, for a flexibility scheme to be successful, it must guarantee that all participants can benefit from participating and are thus willing to engage in the flexibility scheme~\cite{Charalampos_2021}.

Due to driving behaviours, penetration levels, and energy requirements, different EVs add an intrinsic, highly volatile stochasticity layer to the already complex flexibility management problem~\cite{calearo2019grid}. Hence, to successfully implement a flexibility scheme, new management mechanisms are needed that incentivise EV users to offer their flexibility and encourage them to participate in such schemes allowing the DSO to guarantee a high-quality delivery service under uncertainty. In this context, a call for flexibility consists of acquiring services from EVs by the DSO to ensure the safe operation of the grid~\cite{silva2018estimating}, ensuring that EV's interests are respected.



Several works have studied flexibility concepts concerning EVs in distribution systems using pricing strategies. For example, the authors in~\cite{knezovic2017supporting} propose a roadmap with key recommendations for the inclusion of EVs, where they define EV flexibility services in terms of power, time, duration, and location. Furthermore, the authors in~\cite{valogianni2020sustainable} propose an adaptive pricing strategy that helps to mitigate peak demand and to reduce the need for grid reinforcement. Likewise, in \cite{limmer2019peak}, a dynamic pricing strategy for peak load reduction is proposed to optimize the profit of charging pool owners, while the uncertain preferences of customers are accounted for via robust optimization.

Smart charging strategies designed in \cite{canigueral2021flexibility} are able to satisfy multiple flexibility objectives and target specific groups of EV users according to user profile preferences. However, they do not take into account different pricing schemes, aggregator profit, and EV user compensation. Similarly, the authors in \cite{Aragon_2019} present a stochastic optimization model for cooperative control of charging stations using an aggregated energy storage equivalent to describing the charging tasks of the EVs. However, although the approaches mentioned above can provide local peak shaving services, they are not designed to consider network constraints. {In~\cite{Xie_2020}, EV flexibility is provided in the form of peak shaving and valley filling, and pricing and charging scheduling mechanisms are proposed based on a linear demand-price function. The problem is formulated as a bilevel program in which the distribution market clearing is simulated in the lower level  and the upper level solves the EV charging scheduling. Although aggregated flexibility is calculated for DSO services, the proposed framework is deterministic disregarding the uncertain nature of EV parameters.}

The concept of flexibility envelopes was introduced in~\cite{nosair2015flexibility} as an alternative to quantifying flexibility reserves considering the time evolution. This concept has been used, for example, in~\cite{gasser2021predictive}, to show that the flexibility reserves depend highly on the availability of EVs. Furthermore, in \cite{nagpal2022local}, flexibility envelopes are calculated for local energy communities highlighting it as an ease-of-use approach for managing and reserving flexibility in real-time. A similar concept known as \textit{flexibility areas} has been used to estimate the flexibility of the available active and reactive power at the TSO-DSO boundary~\cite{silva2018estimating}. A bottom-up aggregation is commonly performed to estimate such flexibility areas by determining the potential of different assets at the boundary~\cite{fruh2022coordinated}. {In~\cite{Kalantar_2022}, a risk-aware framework is proposed to define the aggregated flexibility from TSO-DSO interconnections and a two-stage linear stochastic optimization model is developed to optimally define the active power flexibility available from DSOs to TSOs via a DC-OPF.} Moreover, as concluded in~\cite{contreras2021computing}, {OPF-based} algorithms allow for obtaining more reliable feasible operating regions compared to random sampling methods. However, none of the above approaches does consider uncertainty.




%
%

Stochastic programming is a common approach for handling uncertainty in electrical power systems including network constraints~\cite{Giraldo_2019}. For example, the authors in~\cite{Alizadeh_2021} introduce a multi-period stochastic \mbox{AC-OPF (SOPF)} considering different flexibility assets for congestion management and voltage control. 
A two-stage stochastic programming model for managing the flexibility of EVs is proposed in~\cite{wu2017two} for distribution systems in which EVs have already been fully recharged. Similarly, the authors in~\cite{sun2020robust} used a linearized power flow model in a stochastic optimization model considering network constraints focusing on the network's reliability. Robust optimization has also been used as in~\cite{arias2021adaptive} to provide flexibility of EVs to DSOs through active and reactive power management strategies minimizing the amount of non-supplied energy and considering network constraints. A queuing network model for electric vehicle charging is presented in \cite{aveklouris2019stochastic}, where the authors define the power allocation in the distribution grid while avoiding congestion and voltage issues. However, all these works assume that users agree to participate in the flexibility scheme without taking into account their particular {priorities}.

{The willingness of participants to engage in energy trading is an essential factor to be considered in a flexibility scheme. Different approaches have been identified in the literature, such as solving a global optimization problem that is aware of all participants' subproblems, double auction schemes, and using marginal utility functions~\cite{Charalampos_2021}. The authors in \cite{sadeghianpourhamami2018quantitive} quantify the EV flexibility for a group of EVs classified by user priorities in terms of amount, time and duration of availability, via a data-driven approach. Even though the EV flexibility is properly quantified, the work focuses on data analysis without explicitly proposing an EV flexibility scheme for practical implementations. Similarly, an online algorithm for charging scheduling of EVs in charging pools is proposed in \cite{mathioudaki2021efficient}, aiming to optimize the amount of energy, charging time and prices for EV users, which are able to choose their most preferable option from a menu-based pricing scheme. Although this work ignores economic profits of each individual charging pool and a detailed operation of the electrical grid (i.e., power flow equations), its online nature sets it as a promising option for real implementations of EV flexibility schemes.} 
 
With this, simplified representations for utility functions are common since they allow for using decentralized optimization algorithms. For example, the authors in~\cite{morstyn2018designing} introduce a decentralized flexibility market based on linear utility functions where the prosumers' willingness to participate is explicitly considered. Furthermore, in~\cite{paudel2020peer} the authors propose using piecewise-quadratic utility functions. However, as found in~\cite{Charalampos_2021}, utility functions are often nonlinear and nonconvex, and in the case they are linear, they can be relatively flat with occasionally significant variations, resulting in non-smooth utility functions.


The reviewed studies show that flexibility services via EV charging have been widely studied. However, we have identified three main gaps in the current literature which we attempt to fill with this paper:   

\begin{itemize}
    \item Most papers dealing with local EV energy management disregard network constraints and do not consider uncertainties. We propose an AC multi-period SOPF considering network constraints and uncertainty related to EV requirements.  
    \item Most papers consider quadratic utility functions because of their attractive properties. We propose a general piecewise-linear formulation that is able to deal with convex and nonconvex utility functions allowing us to represent the interests of EV users. The proposed utility functions represent the participants' willingness to offer flexibility services in the form of energy not served in return for compensation.

    \item {We propose a methodology to estimate risk-aware flexibility areas where the DSO can guarantee operational limits. This is done by introducing a risk parameter representing the willingness of the DSO to withstand operational limit violations. This methodology allows estimating probable costs for flexibility requirements and gives the charging pools more freedom to manage the EV load.}
\end{itemize}

\section{Problem Description}

An operator entity, namely the DSO, is responsible for guaranteeing reliable operational conditions in an electrical distribution network. In addition to constraint satisfaction (i.e., voltage and current magnitude limits), the DSO aims to achieve an economically efficient operation on a day-ahead time frame via flexibility procurement. In this context, we consider a distribution network with a set~$\Omega_{b}$ of nodes, connected by a set of distribution lines. A fixed number of charging pools are connected to the network, identified by the subset $\Omega_{\mathrm{S}}\subset\Omega_{b}$. Hereby, a charging pool $s\in\Omega_{\mathrm{S}}$ consists of a fixed set $\Omega_{\mathrm{N}}^s$ of charging points (e.g., the number of EV parking spaces). A charging task $n$ arriving to the charging pool $s$ is represented as $n\in\Omega_{\mathrm{N}}^s$, and is characterized by its set of requirements $\mathrm{A}_{n}$. It is assumed that a truthful local market mechanism~\cite{Tsaousoglou_2021} is implemented, eliminating any strategic behaviour from the participants, meaning that all charging tasks arriving at a charging pool are willing to provide demand flexibility services in exchange for compensation. This compensation must reflect the charging tasks involved in the process, whether by a tariff reduction, a bonus, or any other kind of settlement~\cite{TSAOUSOGLOU2022}. Therefore, the charging pools act as local flexibility aggregators characterized by a utility function $\boldsymbol{u}_{s}$ which are able to control the charging profiles of their tasks. 

In the implemented market mechanism, the charging pools agree on truthfully communicating the expected requirements of their charging tasks ($\mathrm{A}_{n}$) along with their utility functions ($\boldsymbol{u}_{s}$) to the DSO. Therefore, the DSO aims to obtain optimal demand profiles for the charging pools, which minimize the cost for flexibility procurement while guaranteeing the safe operation of the network over a planning horizon $\Omega_{\mathrm{T}}$. In operation, it would be ideal that the charging pools could provide the demand profiles required, meaning that all operational constraints are satisfied. However, in real operation, the actual delivered power might vary around the planned profiles since the information from the charging pools is intrinsically uncertain, e.g., due to the stochastic behaviour of their charging tasks. Hence, the DSO needs to plan its actions taking into account the operation uncertainties from the charging pools. For this purpose, in this paper, we propose using an AC multi-period stochastic optimal power flow, extending the work in~\cite{Giraldo_2019}.

\begin{figure}[!t]
\centering
  \psfrag{a1}[][][0.8]{{$\textcolor{white}{\boldsymbol{u_{s}}}$}}
  \psfrag{a2}[][][0.8]{{$\textcolor{white}{\mathrm{A}_{n,\omega}}$}}    \psfrag{a3}[][][0.8]{{$\textcolor{new_gray}{\mathcal{Z}_{s}}$}}
\psfrag{a4}[][][0.6]{{$\textcolor{new_gray}{\mathcal{R}_{s,t}}$}} 
  \psfrag{s1}[][][0.6]{{\ref{sec_CT}}}
  \psfrag{s2}[][][0.6]{{\ref{disc_util}}}
  \psfrag{s3}[][][0.6]{{\ref{sopf_model}}}
  \psfrag{s4}[][][0.6]{{\ref{flex_areas_calc}}}
  \psfrag{s5}[][][0.6]{{\ref{sec_5a}}}
  \psfrag{s6}[][][0.6]{{\ref{sec_5c}}}
  \psfrag{s8}[][][0.6]{{\ref{sec_5b}}}

\includegraphics[width=\columnwidth]{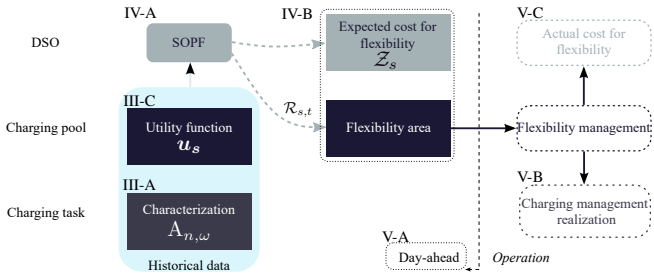}
\caption{{Interaction between DSO, charging pools, and charging tasks.}}
\label{ev_settlement}
\end{figure}

Let $\omega\in\Omega_{\omega}$ be a realization in a set of stochastic scenarios considering possible outcomes due to the uncertainty of the characteristics of the charging tasks. Hence, $\mathrm{A}_{n,\omega}$ represents the expected requirements of charging task $n\in\Omega_{\mathrm{N}}^s$ in scenario $\omega$. The DSO receives this information from the charging pools and solves the SOPF minimizing the expected costs for flexibility $\mathcal{Z}_{s}$ in a day-ahead time frame. {The DSO needs to define a risk parameter $\beta_{s,t}$ based on the risk it is willing to withstand over operational limit violations. Using the optimal solution and the risk parameter, the DSO calculates and communicates a lower and upper power bound to each charging pool valid for each period of the planning horizon. These bounds compose the \mbox{\textit{flexibility area}}, denoted by $\mathcal{R}_{s,t}$.} A graphical representation of the day-ahead planning involving the DSO, charging pools, and charging tasks is depicted in~Fig.~\ref{ev_settlement} {along with its respective section in the paper.} 

In the operation stage, each charging pool $s$ is responsible for the local flexibility management of its charging tasks considering the flexibility area provided by the DSO. This can be done, for example, using profile steering as in~\cite{Gerards_2015}. The actual energy not served to the charging tasks at the end of the day is then aggregated and mapped through the utility function to calculate the actual cost for flexibility.

\section{Mathematical Models} \label{Model}
In this section the different components of the considered setting are presented.
\subsection{Charging tasks} \label{sec_CT}
Consider a charging task $n\in\Omega_{\mathrm{N}}^s$ in charging pool $s\in\Omega_{\mathrm{S}}$ with a maximum deliverable power $\overline{\mathrm{x}}_{n}$. In each scenario $\omega\in\Omega_{\omega}$, a charging task is characterized by the tuple $\mathrm{A}_{n,\omega} =(\mathrm{a}_{n,\omega},\,\mathrm{d}_{n,\omega},\,\mathrm{E}_{n,\omega})$. {The tuple is composed by the task's arrival time $\mathrm{a}_{n,\omega}\in \Omega_{\mathrm{T}}$ following a Poisson distribution characterized by its expected value $\eta_{n}^{\mathrm{a}}$~\cite{aveklouris2019stochastic}. Its departure time $\mathrm{d}_{n,\omega}\in \Omega_{\mathrm{T}}$ as a function of the charging duration following an exponential distribution characterized by the rate $\eta_{n}^{\mathrm{d}}$~\cite{Qian_2011}. And the required charging energy $\mathrm{E}_{n,\omega}$ is assumed to follow a uniform distribution over the closed interval $[\mathrm{e_1,e_2}]$:}
\begin{flalign}
\label{ran_gen}
\mathrm{a}_{n,\omega}\!\sim\!\mathrm{Pois}(\eta_{n}^{\mathrm{a}}),\, \mathrm{d}_{n,\omega}\!\sim\! \mathrm{Exp}(\eta_{n}^{\mathrm{d}}) + \mathrm{a}_{n,\omega},\,\mathrm{E}_{n,\omega}\!\sim\!\mathcal{U}(\mathrm{e_1},\mathrm{e_2})
\end{flalign}
\noindent For feasibility we assume $\mathrm{a}_{n,\omega}< \mathrm{d}_{n,\omega}\leq\left|\Omega_{\mathrm{T}}\right|$, and that within the charging period, the energy required can be delivered at full power, i.e., $\mathrm{E}_{n,\omega} \leq  \left(\mathrm{d}_{n,\omega}-\mathrm{a}_{n,\omega}\right) \overline{\mathrm{x}}_{n}$.

{It is worth mentioning that the effectiveness of the model is independent of the probability distribution function used to model the exogenous stochastic parameters. In fact, these scenarios can also be mapped from real data~\cite{calearo2019grid} or can be synthetically generated~\cite{canigueral2021flexibility, wu2017two, juan_mechanism_2022}}.

\subsection{Charging pools}
A charging pool \mbox{$s \in \Omega_{\mathrm{S}}$}, gets an energy reserve for its charging operation for a future planning horizon $\Omega_{\mathrm{T}}$. The energy reserve is composed of averaged power slots defined before the actual realization $p_{s,t}$~$\forall\,t\in\Omega_{\mathrm{T}}$ and eventual power mismatches $\rho_{s,t,\omega}$ due to the uncertainty of the realizations at each scenario. In other words, $p_{s,t}$ represents the lower power bound of the charging pool at each period, while $\rho_{s,t,\omega}$ represents any consumption above that bound. Let $x_{n,t,\omega}$ be the average power consumption allocated to the charging task $n\in \Omega_{\mathrm{N}}^s$ during timeslot $t$ at the realization of scenario $\omega$. This is a decision variable determined by the charging pool. Then, the power consumption profile of a charging pool $s$ at each stochastic scenario is expressed as:
\begin{flalign} \label{p_station}
&p_{s,t}\!+\!\rho_{s,t,\omega}\!=\!\sum_{n \in \Omega_{\mathrm{N}}^s}x_{n,t,\omega},\,\forall\,s \in \Omega_{\mathrm{S}},\, t \in \Omega_{\mathrm{T}}, \, \omega\in\Omega_{\omega}   
\end{flalign}
\noindent which is limited by an upper bound $\overline{\mathrm{p}}_{s,t}$ representing the power capacity of the charging pool's connection, e.g,. at the transformer
\begin{flalign}
&0 \leq p_{s,t} + \rho_{s,t,\omega} \leq \overline{\mathrm{p}}_{s,t}, & \forall\,s \in \Omega_{\mathrm{S}},\, t \in \Omega_{\mathrm{T}}, \, \omega\in\Omega_{\omega}   
\end{flalign}
\noindent with $p_{s,t},\rho_{s,t,\omega}\geq0$, while the power allocation of each task is bounded by its maximum charging power $\overline{\mathrm{x}}_{n}$:
\begin{multline} \label{bounds}
0 \leq x_{n,t,\omega} \leq \overline{\mathrm{x}}_{n}, \hspace{1pt}  \forall\,s \in \Omega_{\mathrm{S}},\, n \in \Omega_{\mathrm{N}}^s,\, t \in \Omega_{\mathrm{T}}, \, \omega\in\Omega_{\omega}\\:
\mathrm{a}_{n,\omega}\leq t \leq \mathrm{d}_{n,\omega}
\end{multline}
\noindent and power cannot be allocated to task $n$ outside the task's arrival and departure times; hence, \mbox{$ x_{n,t,\omega}=0$} for \mbox{$t<\mathrm{a}_{n,\omega}$ or $\mathrm{d}_{n,\omega}<t$}. Note that vehicle to grid (V2G) can be included by making the left-hand side of~\eqref{bounds} smaller than zero, for example, to allow peer-to-peer transactions inside the charging pool~\cite{arias2021adaptive}.   
%

The charging pools also offer flexibility which may imply that some charging tasks end with a lower charged energy than initially requested. This leads to energy not served at task $n$ in scenario $\omega$, defined as $\phi_{n,\omega}$:
\begin{multline} \label{serve}
\mathrm{E}_{n,\omega} = \Big( \sum_{\mathclap{ \substack{t\in \Omega_{\mathrm{T}}} } }\,x_{n,t,\omega}\Big)\!+\!\phi_{n,\omega},\\ \forall\,s \in \Omega_{\mathrm{S}},\, n \in \Omega_{\mathrm{N}}^s , \, \omega\in\Omega_{\omega}.    
\end{multline}
%

For charging pool $s$, the total amount of energy not served to its charging tasks is expressed as:
\begin{flalign} \label{non_served_aggre}
&\Phi_{s,\omega} = \sum_{\mathclap{n \in \Omega_{\mathrm{N}}^s}}\,\phi_{n,\omega} , &\forall\,s \in \Omega_{\mathrm{S}}, \, \omega\in\Omega_{\omega}.    
\end{flalign}



\subsection{Discrete utility functions} \label{disc_util}

The utility function $\boldsymbol{u}_{s}$ of a charging pool \mbox{$s \in \Omega_{\mathrm{S}}$} expresses the cost for flexibility as a function of the total energy not served $\Phi_{s,\omega}$. {It has been recognized that actual utility functions can be highly nonlinear~\cite{Charalampos_2021} and also not necessarily convex. For this reason, a general formulation is needed to approximate any realistic utility function. To this end,} we propose the use of discretized utility functions using a semicontinuous convex combination formulation~\cite{vielma2008nonconvex}. {This formulation does not rely on the nature of the utility function (monotonicity or convexity) to approximate it.}

\begin{figure}[t]
\centering
  \psfrag{u}[][][0.9]{{$\boldsymbol{u_{s}}$}}
   \psfrag{a0}[][][0.8]{{$\alpha_{s,0}$}}
 \psfrag{a1}[][][0.8]{{$\alpha_{s,1}$}}
  \psfrag{a2}[][][0.8]{{$\alpha_{s,2}$}}
  \psfrag{a3}[][][0.8]{{$\alpha_{s,3}$}}
  \psfrag{phi}[][][0.85]{{$\boldsymbol{\Phi_{s,\omega}}$}}
  \psfrag{f}[][][0.8]{{$f_{s,2}$}}
    \psfrag{f1}[][][0.8]{{$f_{s,1}$}}
  \psfrag{f3}[][][0.8]{{$f_{s,3}$}}

  \psfrag{b1}[][][0.8]{{$\overline{\mathrm{u}}_{s,1}$}}
  \psfrag{b2}[][][0.8]{{$\underline{\mathrm{u}}_{s,2}$}}
\includegraphics[width=0.8\columnwidth]{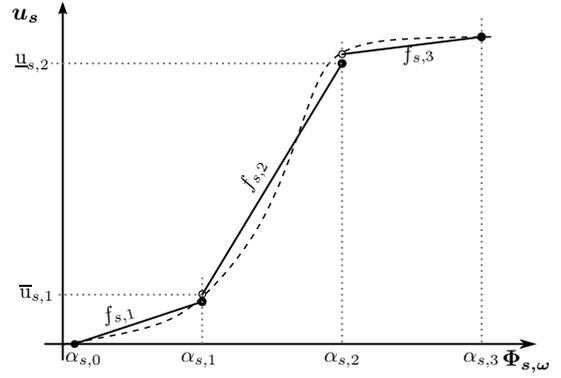}
\caption{Representation of a utility function for a charging pool $s$ with $\kappa=3$.}
\label{util_disc}
\end{figure}

An example of a utility function $\boldsymbol{u}_{s}$ is shown in Fig.~\ref{util_disc}, where the dashed line represents a continuous nonlinear function, approximated by a linear piecewise function with three segments. 

In this work we consider a lower-semicontinuous piecewise-linear function representing the utility function of the charging pool $s$:
\begin{flalign} \label{util_funct_original}
&\boldsymbol{u}_{s}\!=\!\begin{cases}
      f_{s,0}\!=\!0 & \Phi_{s,\omega} = 0 \\
      f_{s,1} = \mathrm{h}_{s,1}\Phi_{s,\omega}\!+\!\mathrm{b}_{s,1} & 0 < \Phi_{s,\omega} \leq \alpha_{s,1} \\
      \vdots \\ 
      f_{s,\kappa}\!=\!\mathrm{h}_{s,\kappa}\Phi_{s,\omega}\!+\!\mathrm{b}_{s,\kappa} & \alpha_{s,\kappa-1}\!<\!\Phi_{s,\omega}\!\leq \alpha_{s,\kappa}
   \end{cases}
\end{flalign}

\noindent  where $k\in\Omega_{\mathrm{K}}^{s}=\left\{0,\,\ldots,\,\kappa\right\}$ represents the set of break points, while $\{\mathrm{h}_{s,k},\,\mathrm{b}_{s,k}\}$ and $\{\alpha_{s,k-1},\,\alpha_{s,k}\}$ $\forall\,k\geq1$ denote the coefficients of the functions and their lower and upper bounds, respectively. For the sake of simplicity, we take \mbox{$\overline{\mathrm{u}}_{s,k-1}:= f_{s,k}\left(\alpha_{s,k-1}\right)$} and \mbox{$\underline{\mathrm{u}}_{s,k}:= f_{s,k}\left(\alpha_{s,k}\right)$} as the function of segment $k:k\geq1$ evaluated on its endpoints. In order to satisfy~\eqref{util_funct_original}, notice that \mbox{$\underline{\mathrm{u}}_{s,0}=\overline{\mathrm{u}}_{s,0}=\alpha_{s,0}=0$}. 

It must be pointed out that~\eqref{util_funct_original} cannot be directly integrated into a mathematical programming model. However, by defining multipliers $\overline{\lambda}_{s,k,\omega}$ $\underline{\lambda}_{s,k,\omega}\geq 0\; \forall\,k\in\Omega_{\mathrm{K}}^{s}$ as the weights at each two endpoints, and binary variables $y_{s,k,\omega}$, the utility function can be expressed as a linear combination of the cost of the endpoints by:      
%
\begin{flalign}
\label{util_funct_piece}
\mathcal{Z}_{s,\omega} = \sum_{\mathclap{{k\in\Omega_{\mathrm{K}}^{s}}:\,k<\kappa}}\left(\underline{\lambda}_{s,k,\omega}\underline{\mathrm{u}}_{s,k}+\overline{\lambda}_{s,k,\omega}\overline{\mathrm{u}}_{s,k}\right) + \underline{\lambda}_{s,\kappa,\omega}\underline{\mathrm{u}}_{s,\kappa} 
\end{flalign}
%
\noindent where the energy not supplied is defined as:
%
\begin{multline}
\label{non_served_rep}
\Phi_{s,\omega}\!=\!\sum_{\mathclap{{k\in\Omega_{\mathrm{K}}^{s}}:\,k<\kappa}}\left(\underline{\lambda}_{s,k,\omega}\!+\!\overline{\lambda}_{s,k,\omega}\right)\alpha_{s,k}+\!\underline{\lambda}_{s,\kappa,\omega}\alpha_{s,\kappa} \\ 
\forall\,s \in \Omega_{\mathrm{S}},\, \omega \in \Omega_{\omega}
\end{multline}

To make sure that~\eqref{util_funct_piece} and~\eqref{non_served_rep} lead to a proper representation of the utility function, the following constraints are added:
\begin{flalign}\label{utili3}
&1\!=\!\sum_{\mathclap{{k\in\Omega_{\mathrm{K}}^{s}}:\,k<\kappa}}\left(\!\underline{\lambda}_{s,k,\omega}\!+\!\overline{\lambda}_{s,k,\omega}\!\right) \!+\!\underline{\lambda}_{s,\kappa,\omega}, \hspace{1pt} \forall\,s \in \Omega_{\mathrm{S}},\, \omega \in \Omega_{\omega}
\end{flalign}
%
%
%
\begin{multline}\label{utili1}
\overline{\lambda}_{s,k,\omega}\!+\!\underline{\lambda}_{s,k+1,\omega}\!=\!y_{s,k+1,\omega} \\ \forall s \in \Omega_{\mathrm{S}}, k\in\Omega_{\mathrm{K}}^{s},\, \omega \in \Omega_{\omega}:\,k<\kappa
\end{multline}
\vspace{-10pt}
\begin{flalign}\label{utili2}
&\sum_{{{k\in\Omega_{\mathrm{K}}^{s}}:k\geq1}} y_{s,k,\omega} \leq 1 & \forall \,s \in \Omega_{\mathrm{S}},\, \omega \in \Omega_{\omega}
\end{flalign}
\vspace{-10pt}
\begin{flalign}
\label{utili22}
& y_{s,k,\omega}\in\{0,1\} &\forall s \in \Omega_{\mathrm{S}},\, k\in\Omega_{\mathrm{K}}^{s},\, \omega \in \Omega_{\omega}:k\geq1
\end{flalign}

Notice that~\eqref{utili3} and \eqref{utili1} ensure that the multipliers are only different from zero in the segment where $y_{s,k,\omega}$ is activated, while~\eqref{utili2} and~\eqref{utili22} guarantee that only one segment can be active. Hence, considering the utility functions, the set of variables from the charging pools is defined as \mbox{$\mathcal{Y}_{cp}=\{\Phi_{s,\omega},\mathcal{Z}_{s,\omega}, \overline{\lambda}_{s,k,\omega},\underline{\lambda}_{s,k,\omega}, y_{s,k,\omega},x_{n,t,\omega}\}$}.   

\subsection{Distribution Network Model}
 %

We consider a distribution network with radial topology behind an electrical substation denoted by ES and a set of branches~$\Omega_{l}\subset\Omega_{b}\times\Omega_{b}$. The operational state of the network for a given scenario $\omega\in\Omega_{\omega}$ can be calculated based on the power flow equations as given in constraints~$\eqref{P_balance}\textrm{--}\eqref{limit_corri}$, adapted from~\cite{Giraldo_2017}. 
Hereby, the active power balance in the network is ensured by:
%
\begin{multline}
\label{P_balance}
\sum_{\mathclap{m i \in \Omega_{l}}} P_{m i,t,\omega}-\sum_{\mathclap{ij \in \Omega_{l}}} \left(P_{ij,t,\omega} + \mathrm{R}_{ij} I_{ij,t,\omega}^{\mathrm{sqr}} \right)+P^{\mathrm{G}}_{i,t,\omega}= \mathrm{P}_{i,t}^{\mathrm{D}} + \\ + \sum_{\mathclap{s\in \Omega_{\mathrm{S}} : s=i}}p_{s,t} + \rho_{s,t,\omega} \hspace{20pt} \forall\,i \in \Omega_{b},\, t \in \Omega_{\mathrm{T}},\, \omega \in \Omega_{\omega}
\end{multline}
and the reactive power balance is given by:
%
\begin{multline} \label{Q_balance}
\sum_{\mathclap{m i \in \Omega_{l}}} Q_{m i,t,\omega} - \sum_{\mathclap{ij \in \Omega_{l}}} \left(Q_{ij,t,\omega} + \mathrm{X}_{ij} I_{ij,t,\omega}^{\mathrm{sqr}} \right)+Q^{\mathrm{G}}_{i,t,\omega}=\mathrm{Q}_{i,t}^{\mathrm{D}},\\ \forall i \in \Omega_{b}, t \in \Omega_{\mathrm{T}},\, \omega \in \Omega_{\omega}
\end{multline}
where $\mathrm{P}_{i,t}^{\mathrm{D}}$ and $\mathrm{Q}_{i,t}^{\mathrm{D}}$ denote the {regular} active and reactive power demands at node $i$ and timeslot $t$. {Regular} power demands are assumed to be deterministic parameters expressing the base load of all nodes disregarding EVs. Doing this allows us to focus on the impact of EVs. Active and reactive power flows to node $i$ from its parent node $m$ are denoted by $P_{m i,t,\omega}, Q_{m i,t,\omega}$, while $P_{ij,t,\omega}, Q_{ij,t,\omega}$ are the active/reactive power flows from node $i$ to its descendant nodes~$j$. For the purposes of this work, it is also assumed that the charging stations operate at a unitary power factor and no other controllable power sources, such as distributed generators are available in the network, hence \mbox{$P^{\mathrm{G}}_{i,t,\omega} = Q^{\mathrm{G}}_{i,t,\omega} = 0, \forall i \in \Omega_b:i\neq \mathrm{ES}$}. Also, the voltage magnitude is assumed to be known for the substation \mbox{($V_{\mathrm{ES},t,\omega}^{\mathrm{sqr}} = 1.0$~pu)}. The voltage magnitude drop between nodes $i$ and $j$ is represented by:
\begin{multline} \label{eqn:tensi}
V_{j,t,\omega}^{\mathrm{sqr}} = V_{i,t,\omega}^{\mathrm{sqr}}-2\left(\mathrm{R}_{ij} P_{ij,t,\omega}\!+\!\mathrm{X}_{ij} Q_{ij,t,\omega}\right)\!+ \\  - \left(\mathrm{R}_{ij}^2\!+\!\mathrm{X}_{ij}^2\right) I_{ij,t,\omega}^{\mathrm{sqr}}, \hspace{10pt}
\forall ij \in \Omega_{l}, t \in \Omega_{\mathrm{T}},\, \omega \in \Omega_{\omega}
\end{multline}
\noindent where $V_{i,t,\omega}^{\mathrm{sqr}}:=V_{i,t,\omega}^{2}$ and $I_{ij,t,\omega}^{\mathrm{sqr}}:=I_{ij,t,\omega}^{2}$ are defined to obtain a convex relaxation of the problem~\cite{gan2014exact}, while branch power flows are obtained using the rotated second-order cone constraint:
\begin{multline}\label{eqn:corri}
V_{j,t,\omega}^{\textrm{sqr}} I_{ij,t,\omega}^{\textrm{sqr}} \geq P_{ij,t,\omega}^{2} + Q_{ij,t,\omega}^{2}, \\\forall ij \in \Omega_{l}, t \in \Omega_{\textrm{T}},\, \omega \in \Omega_{\omega}.
\end{multline} 

Furthermore, the upper and lower bounds for nodal voltage and branch current magnitudes are enforced by
\begin{flalign}\label{lim_tensi} 
&\underline{\textrm{V}}^2 \leq V_{i,t,\omega}^{\textrm{sqr}} \leq \overline{\textrm{V}}^2  & \forall i \in \Omega_{b}, t \in \Omega_{\textrm{T}},\, \omega \in \Omega_{\omega} \\
\label{limit_corri}
& 0 \leq I_{ij,t,\omega}^{\textrm{sqr}} \leq \overline{\textrm{I}}_{ij}^{2} &\forall ij \in \Omega_{l}, t\in \Omega_{\textrm{T}},\, \omega \in \Omega_{\omega}
\end{flalign}

Finally, the set of variables from the distribution network is denoted by $\mathcal{Y}_{dn}=\{V_{i,t,\omega}^{\mathrm{sqr}},I_{ij,t,\omega}^{\mathrm{sqr}},P_{ij,t,\omega}, Q_{ij,t,\omega},\rho_{s,t,\omega}\}$. 

\section{Proposed SOPF Model and Estimation of Flexibility Areas}\label{prop_meth}

\subsection{Mathematical model} \label{sopf_model}

Using the mathematical formulations given in the previous section, the proposed SOPF is cast as a two-stage stochastic optimization model, formulated as:
%
\begin{flalign}
\begin{split}
\label{opf_nonconvex}
&\min_{\boldsymbol{\mathcal{Y}}}\hspace{5pt}\sum_{\omega \in \Omega_{\omega}}\pi_{\omega}\sum_{s \in \Omega_{\mathrm{S}}}\mathcal{Z}_{s,\omega} - \sum_{t \in \Omega_{\mathrm{T}}}\sum_{s \in \Omega_{\mathrm{S}}} \mathrm{c}_{s,t}\,p_{s,t}\\
&\mathrm{\mathrm{s.t.}}
\hspace{20pt} \eqref{p_station}\textrm{--}\eqref{non_served_aggre} ,\, \eqref{util_funct_piece}\textrm{--}\eqref{limit_corri}
\end{split}
\end{flalign} 
\noindent where the set \mbox{$\boldsymbol{\mathcal{Y}}=\{\mathcal{Y}_{cp}\cup \mathcal{Y}_{dn}\cup p_{s,t}\}$} contains the decision variables of the model, $\pi_{\omega}$ stands for the probability of scenario $\omega$, and $\mathrm{c}_{s,t}$ represents the unit cost of electricity at charging pool $s$ at time~$t$. The first-stage variables (\mbox{\textit{here-and-now}}) are $p_{s,t}$, representing the decisions the DSO takes in advance without knowing the actual realizations, while the second-stage variables (\mbox{\textit{wait-and-see}}) are $\mathcal{Y}_{cp}$ and $\mathcal{Y}_{dn}$, representing the expected stochastic behavior of the system after fixing the first-stage variables. 
Note that although DSOs are not allowed to retail electricity, they may procure flexibility from the charging pools, which act as local flexibility aggregators. Hence, the objective function in~\eqref{opf_nonconvex} minimizes the expected value of the cost for flexibility and maximizes the energy reserved for the charging pools.  

It must be pointed out that the SOPF in~\eqref{opf_nonconvex} is based on a mixed-integer second-order cone programming \mbox{(MISOCP)} problem, which is nonconvex in principle. However, if the two sufficient conditions defined in~\cite{gan2014exact} are satisfied, then the relaxed continuous equivalent is convex and exact, and a globally optimal solution is numerically reachable~\cite{bonami2008algorithmic}. In the presented model, both conditions are satisfied since the only power source in the system is the substation. Thus, every node only consumes power, and the upper bounds of the voltages are not binding as long as $V_{\mathrm{ES},t}<\overline{\textrm{V}}$. Moreover, a numerical solution to~\eqref{opf_nonconvex} can be obtained using the sample average approximation (SAA) technique under different scenario generation methods, e.g., Monte Carlo (MC), moment matching, or point estimate methods~\cite{Giraldo_2019}. 

\subsection{Estimation of the flexibility areas}\label{flex_areas_calc}

Based on the optimal solution $\boldsymbol{\mathcal{Y}^{*}}$ of the SOPF in~\eqref{opf_nonconvex}, the empirical cumulative density function (eCDF) of $\rho_{s,t,\omega}^{*}$ can be calculated, which is denoted as $F_{\rho_{s,t}}$. Hence, the flexibility area of a charging pool $s$ at period $t$ is calculated as
\begin{flalign} \label{op_room_eq}
\mathcal{R}_{s,t} = p_{s,t}^{*} + F^{-1}_{\rho_{s,t}}\left(\beta_{s,t}\right).
\end{flalign}

\noindent where \mbox{$\beta_{s,t} \in [0,1]$} represents a risk parameter defined by the DSO for each charging pool at each time period. Notice that the flexibility area $\mathcal{R}_{s,t}$ is composed of two terms, the power reserve serving as a lower limit ($p_{s,t}^{*}$) and the upper limit calculated for a specified quantile. {It is worth noting that the risk of violating the operational limits and the flexibility area are directly proportional.} {This means that \mbox{$\beta_{s,t}=0$} represents the most conservative alternative (lowest risk/smallest area), i.e., \mbox{$\mathcal{R}_{s,t}=p_{s,t}^{*}$}, while the most optimistic alternative (highest risk/biggest area) is given for \mbox{$\beta_{s,t}=1$}, leading to \mbox{$\mathcal{R}_{s,t}=p_{s,t}^{*}+\max\limits_{\omega}\{\rho_{s,t,\omega}^{*}\}$}.} 

Furthermore, from the perspective of the charging pools, the flexibility area can be interpreted as an accepted operating region to fulfill its charging duties within which the DSO expects to guarantee operational limits. {Finally, notice that $\mathcal{R}_{s,t}$ can only be obtained after solving~\eqref{opf_nonconvex} since it depends on the optimal solution to uncertain realizations. }   %

\section{Test System and Simulations} \label{sec_5}

\begin{figure}[!tb]
\centering
 \psfrag{lb}[][][0.5]{{{Regular} demand}}
  \psfrag{cs}[][][0.5]{{Charging pool}}
    \psfrag{ts}[][][0.5]{{Charging task}}
\includegraphics[width=0.9\columnwidth]{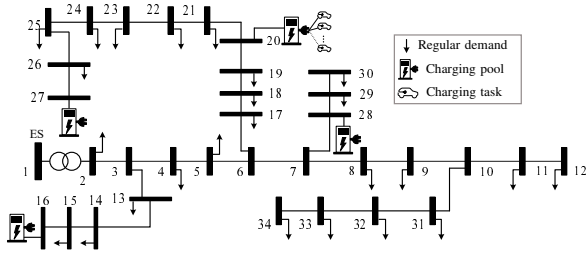}
\caption{34-nodes test system including four charging pools.}
\label{system_fig}
\end{figure}

In this section, we evaluate the proposed stochastic flexibility model. For the tests, we consider a radial distribution system modified from~\cite{Giraldo_2017} with 34 nodes (see Fig.~\ref{system_fig}), which is an 11~kV network with a peak total nominal power of 1.86\,MW, 1.23\,Mvar, $\underline{\mathrm{V}}=0.95$\,pu, and $\overline{\mathrm{V}}=1.05$\,pu. The maximum phase current at the substation transformer connecting nodes 1-2 has been set to $\overline{\mathrm{I}}_{1\,2}=88$~A. Four charging pools are placed at nodes 16, 20, 27, and 28, with 30, 59, 36, and 16 charging tasks spread over the planning horizon, respectively. The planning horizon is discretized in $24$ one-hour intervals, resembling a day-ahead planning procedure.

\begin{figure}[t]
\centering
\includegraphics[width=\columnwidth]{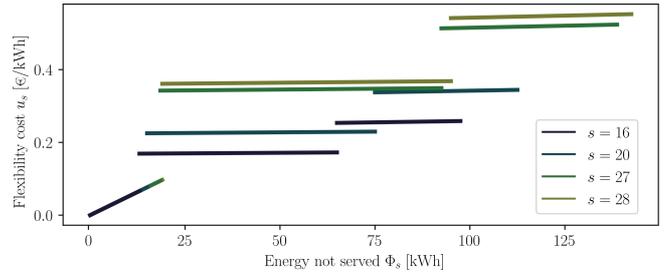}
\caption{Utility functions used by the charging pools.}
\label{util_fig}
\end{figure}

The shape parameters $\eta_{n}^{\mathrm{a}},\,\eta_{n}^{\mathrm{d}}$, characterizing the arrival and duration times for the EV charging tasks, were obtained considering the data in~\cite{canigueral2021flexibility} for weekdays. A Monte Carlo SAA with $\left|\Omega_{\omega}\right|=500$ was used to solve the two-stage SOPF~\eqref{opf_nonconvex}, considering equiprobable scenarios, i.e., $\pi_{\omega} = 1/\left|\Omega_{\omega}\right|$.  The arrival and departure times for each scenario were calculated as in~\eqref{ran_gen}, while the energy required at each scenario was calculated as \mbox{$\mathrm{E}_{n,\omega}=\min\{\mathcal{U}(\mathrm{e_1},\mathrm{e_2}),\,\overline{\mathrm{x}}_{n}\left(\mathrm{d}_{n,\omega}-\mathrm{a}_{n,\omega}\right)\}$} with \mbox{$\mathrm{e}_{1} = 0$\,kWh and $\mathrm{e}_{2} = 100$\,kWh}.  Without loss of generality, the maximum power at each charging pool has been set to $\overline{\mathrm{p}}_{s,t}=200$\,kW, a fixed cost for electricity of $\mathrm{c}_{s,t} = 0.2$~\texteuro/kWh was chosen, and the maximum power at each charging task was set to $\overline{\mathrm{x}}_{n} = 22$\,kW. Finally, the utility functions for the four charging pools have been parameterized as in Fig.~\ref{util_fig} with~$\kappa=3$. 

\subsection{Obtaining Flexibility Areas -- Day-ahead planning} \label{sec_5a}

\begin{figure}[t]
\centering
\includegraphics[width=\columnwidth]{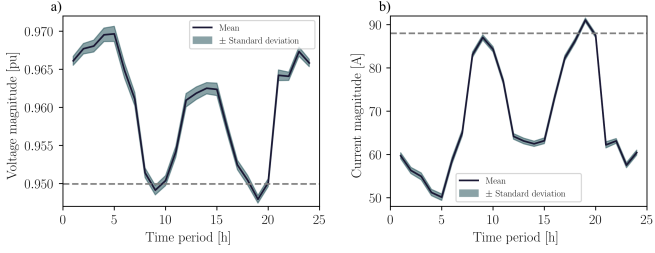}
\caption{Base case results for the planning horizon indicating congestion problems. (a)~Lowest voltage magnitude. (b)~Highest current magnitude.}
\label{V_mag_27}
\end{figure}

Two main tests were carried out to determine the flexibility areas. The first one corresponds to the base case, an instance with relaxed voltage and current magnitude constraints and disabled flexibility from charging pools. The base case corresponds to a situation where all required energy from charging tasks is supplied as soon as possible, regardless of the network status. The mean and standard deviation of the minimum voltage magnitude at each time period and the maximum branch current at each time period for the base case are shown in Fig.~\ref{V_mag_27}. In Fig.~\ref{V_mag_27}~(a), periods with undervoltage problems can be seen in around 8-10h and 18-20h. Similarly, periods with overloading problems are evident in Fig.~\ref{V_mag_27}~(b) around 18-20h. These results indicate that the DSO might have a congestion problem during the planning horizon and the need for flexibility.

The second test corresponds to the opposite case, i.e., operational constraints are enforced and flexibility from charging pools is enabled. The resulting value of the objective function found was \mbox{(-)\texteuro$4{,}059.12$} for the base case and \mbox{(-)\texteuro$3{,}908.78$} in the flexibility enabled case. Notice that lower values indicate less energy not served. These results represent a reduction of 3.7\% in the total expected payment due to the flexibility cost in the latter case. These results indicate that the charging pools (aggregators) would need to pay for the energy not served to some charging tasks to comply with the DSO's expected flexibility requirements. Consequently, it is expected that the DSO settles this difference with the charging pools as part of a flexibility market~\cite{TSAOUSOGLOU2022}.


%
\begin{figure}[!b]
\centering
\includegraphics[width=\columnwidth]{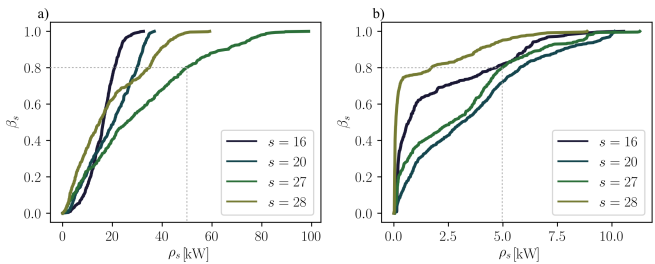}
\caption{eCDFs of the operational power $\rho_{s,t,\omega}$ for (a) $t=14$h and (b) $t=19$h.}
\label{cdf_pdf_rho}
\end{figure}

\begin{figure}[!t]
\centering
\includegraphics[width=\columnwidth]{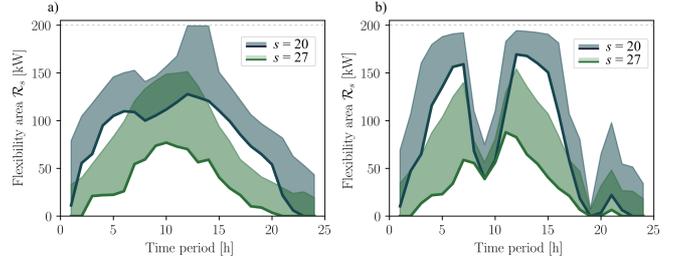}
\caption{Flexibility area for $\beta_{s,t} = 0.9$ at charging pools $s=20$ and $s=27$. \mbox{(a) Base case.} (b) Flexibility enabled.}
\label{op_room}
\end{figure}

The flexibility areas proposed in Section~\ref{flex_areas_calc} allow the DSO to estimate safe operation regions for the charging pools. The first step to obtain the flexibility areas is calculating the empirical eCDF of $\rho_{s,t,\omega}^{*}$ based on~\eqref{op_room_eq}. The eCDF of the four charging pools at $s=\{16,\,20,\,27,\,28\}$ are shown in Fig.~\ref{cdf_pdf_rho}(a) for $t=14$h and in Fig.~\ref{cdf_pdf_rho}(b) for $t=19$h. It can be seen that the expected power areas chosen depend on the period, e.g., for $\beta_{27}=0.8$ the operational powers need to be lower than or equal to \mbox{$\rho_{27}=49.61$~kW} at $t=14$, but lower than or equal to $\rho_{27}=4.96$~kW at $t=19$. This difference is expected due to the network's characteristics, i.e., there are some periods where the charging pools can have more \textit{room} to supply their charging tasks without compromising the network's operational limits than in other periods. The flexibility area, which finally will be communicated to the charging pools, has been calculated using~\eqref{op_room_eq} for both test cases. In~\eqref{op_room_eq}, the flexibility area is composed by two terms, the power reserve serving as a lower limit (bold line) and the upper limit calculated for a specified quantile, as shown in Fig.~\ref{op_room} for $s=20$ and $s=27$ using $\beta_{s,t} = 0.9$. The load shifting is evident when comparing both test cases during the whole time horizon, especially during critical time intervals (8-10h and 18-20h). 

\begin{figure}[!t]
\centering
\includegraphics[width=\columnwidth]{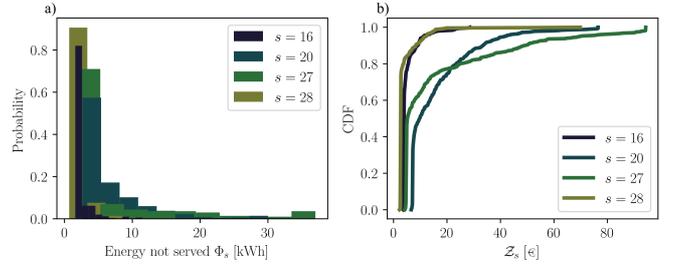}
\caption{(a) PDFs of the total energy not served. (b) eCDFs of the total cost for flexibility.}
\label{cdf_pdf}
\end{figure}

However, load shifting is not always sufficient to solve the congestion problems in this test case. Therefore, the charging pools must also procure flexibility from the charging tasks in the form of energy not served to guarantee the operational limits of the DSO. The probability density function (PDF) of the total energy not served at the four charging pools is displayed in Fig.~\ref{cdf_pdf} (a). Similarly, Fig.~\ref{cdf_pdf}~(b) presents the eCDF of the cost for flexibility at each charging pool. It can be seen that the most procured charging pools are $s=20$ and $s=27$, which belong to the same network feeder (see Fig.~\ref{system_fig}).
Interestingly, for this feeder, the most pronounced voltage drops occur; hence, the DSO must procure flexibility in these two charging pools to solve voltage problems. It is then evident that some charging pools can have an advantageous market position and might behave strategically depending on their location in the network (e.g., due to the radial topology of distribution networks). Therefore, these results reinforce the importance of truthful and fair market mechanisms in future flexibility markets~\cite{Tsaousoglou_2021,TSAOUSOGLOU2022}.  

Moreover, from Fig.~\ref{cdf_pdf} (b), the DSO can estimate the expected cost for flexibility at each charging pool. For example, using the 90th percentile for $s=27$, means that the cost for flexibility at that charging pool is expected to be lower than or equal to \texteuro~48.97 in at least 90\% of the expected scenarios.

\subsection{Validation of the Obtained Flexibility Areas with Probabilistic Power Flow -- Operation} \label{sec_5c}

The next step considers an operation scenario based on the flexibility areas identified for day-ahead in Sec.~\ref{sec_5a}. Two risk values are tested in this section to show the impact of $\beta_{s,t}$ on the safe operation of the system. We took arbitrarily risk values $\beta_{s,t}\in\{0.57,\,0.99\}$ for the following analysis. A probabilistic power flow consisting of $5{,}000$ MC simulations is executed, considering the uncertainties of the aggregated consumed power at the charging pools. A sequential implementation of the power flow given in~\cite{giraldo2022fixed} has been used due to its convergence and computational characteristics. Uniform distributions are assumed to cope with any scenario combination within the flexibility area defined by the selected risk value of the form
$\sim \mathcal{U}(p_{s,t},\mathcal{R}_{s,t})$. It is assumed that the charging pools are able to control their consumption within the required flexibility area. Finally, it must be pointed out that voltage and current magnitude limits are not enforced in the power flow.  

\begin{figure}[!t]
\centering
\includegraphics[width=\columnwidth]{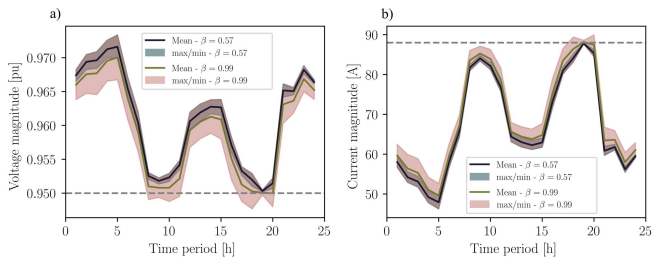}
\caption{CDF of operation results  during . (a)~Lowest voltage magnitude. (b)~Highest current magnitude.}
\label{operation_VI}
\end{figure}

At each MC simulation, the lowest voltage and the highest current magnitudes of the system per time period are stored. In Fig.~\ref{operation_VI}(a), the average of the lowest voltage magnitude among the buses using both risk values is the continuous line, while the shaded area indicates its maximum and minimum values. Similarly, Fig.~\ref{operation_VI}(b) displays the average maximum branch current magnitude and its maximum and minimum values. For instance, at 20\,h the average lowest voltage magnitude for $\beta_{s,t}=0.57$ is $0.9514$\,pu with a maximum of $0.9522$\,pu and a minimum of $0.9507$\,pu. The maximum current magnitude at the same time has an average of $85.37$\,A, a maximum of $86.27$\,A and a minimum of $84.48$\,A. On the other hand, for $\beta_{s,t}=0.99$, the average lowest voltage magnitude is $0.9499$\,pu with a maximum of $0.9521$\,pu and a minimum of $0.9479$\,pu; while the current magnitude has an average of $87.43$\,A, a maximum of $89.95$\,A and a minimum of $84.69$\,A.

\begin{figure}[!t]
\centering
\includegraphics[width=\columnwidth]{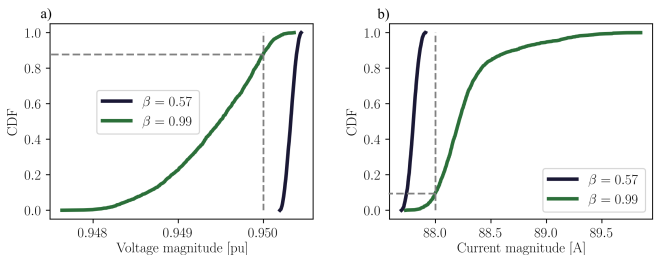}
\caption{Operation results for the planning horizon using different flexibility areas. (a)~Lowest voltage magnitude. (b)~Highest current magnitude.}
\label{operation_VI_cdf}
\end{figure}

The eCDFs of the minimum voltage magnitude considering all periods is depicted in Fig.~\ref{operation_VI_cdf}(a) for both risk values. It can be seen that around $88\%$ of the scenarios violate the voltage limit for $\beta_{s,t}=0.99$, whereas for $\beta_{s,t}=0.57$ minimum voltages are always within the limit. The eCDFs of the maximum current magnitudes are displayed in Fig.~\ref{operation_VI_cdf}(b) where a similar result is obtained with only $10\%$ of the scenarios respecting the maximum current magnitude limit when $\beta_{s,t}=0.99$. These results indicate that the DSO must determine the required flexibility areas based on the risk it is willing to accept since there is a trade-off between the chosen risk value and the probability of violating the operational limits. 

\subsection{Impact of Flexibility Areas on the Total Payment of the Charging Pools} \label{sec_5b}

\begin{figure}[!t]
\centering
\includegraphics[width=\columnwidth]{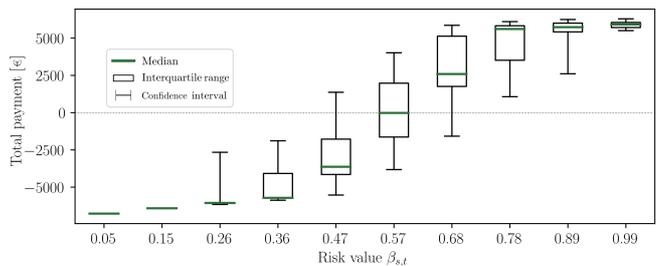}
\caption{Total payment from charging pools for different risk values.}
\label{boxplot_payment}
\end{figure}

A final test is performed to assess the impact of the flexibility areas on the total payment received by the charging pools. We considered ten risk values used by the DSO (see Fig.~\ref{boxplot_payment} for the chosen values). The obtained flexibility areas for the different risk values were taken as power limiters for the charging pools, i.e., $\overline{\mathrm{p}}_{s,t} = \mathcal{R}_{s,t}$. On the other hand, the total payment, representing the revenue of the charging pools, was calculated as the difference between the cost for the energy delivered to their charging tasks and the cost for energy not served. Thus, positive total payment values are desired to guarantee revenue adequacy~\cite{ming2018revenue}. We simulated 1{,}000 random scenarios for each risk value, following the same distributions as described earlier for the random variables. Voltage and current magnitude limits were enforced and the flexibility enabled.

The obtained results are displayed in Fig.~\ref{boxplot_payment} using a box plot where the median, the interquartile range, and the 90\% confidence intervals are depicted. Results for $\beta_{s,t}=0.57$ show that the median is \mbox{\texteuro\,-18.48}, the interquartile range is limited by \texteuro\,1{,}978.87 and \texteuro\,-1{,}635.18, and the confidence interval is \texteuro\,4{,}021.13 and \texteuro\,-3{,}826.80; whereas for $\beta_{s,t}=0.99$ all these values increased considerably. Hence, it can be seen that the total payment for flexibility increases with the risk value, meaning there is a trade-off between the risk the DSO is willing to stand and the revenue of the charging pools. Interestingly, risk values $\beta_{s,t} <0.57$ might produce revenue inadequate situations, which encourages the use of proper compensation mechanisms for energy not served~\cite{juan_mechanism_2022}. {Consequently, it is expected that the DSO settles this difference with the charging pools as part of a flexibility market~\cite{TSAOUSOGLOU2022}.           }

\section{Conclusions}
In this paper, we proposed a stochastic AC-OPF for the flexibility management of charging pools in distribution networks {introducing the concept of \textit{flexibility areas}}. The SOPF considers discrete utility functions for charging pools as a compensation mechanism for eventual energy not served to their charging tasks. The utility functions are presented using a general piecewise-linear formulation to deal with convex and nonconvex prosumer preferences. The aim is to minimize the expected cost for energy not served while satisfying operational constraints.
An application of the proposed SOPF has been described, where a DSO specifies the flexibility area to each charging pool in a day-ahead time frame under uncertainty. {This methodology allows estimating probable costs for flexibility requirements and gives the charging pools more freedom to manage the EV load.} Results show that a safe flexibility area for charging pools can be used to address DSO's congestion problems, either by load shifting or managing the energy not served. Moreover, the DSO is able to calculate the flexibility area as a function of a risk parameter $\beta_{s}$ and estimate probable costs for flexibility requirements. Results showed a trade-off between the risk the DSO is willing to stand and the revenue of the charging pools. At the same time, charging pools and tasks perceive a total energy payment reduction as compensation for the energy not served, which might stimulate charging pool operators and EV users to offer flexibility services (e.g., in a local flexibility market). Future work has to analyze the impact of the proposed flexibility area considering V2G enabled EVs and reactive power compensation capabilities.




\bibliographystyle{IEEEtran}
\bibliography{references}

\begin{IEEEbiographynophoto}{Juan S. Giraldo} received the B.Sc. degree in electrical engineering from the Universidad Tecnológica de Pereira, Pereira, Colombia, in 2012, and the M.Sc. and Ph.D. degrees in electrical engineering from the University of Campinas (UNICAMP), Campinas, Brazil, in 2015 and 2019, respectively. From Oct. 2019 to May 2021 he was a Postdoctoral Fellow at the Department of Electrical Engineering, Eindhoven University of Technology, Eindhoven, The Netherlands (NL). Later, from June 2021 to Aug. 2022 he was a postdoc with the Mathematics of Operations Research group at the University of Twente, Enschede, NL. He is currently a Researcher with the Energy Transition Studies group with the Netherlands Organisation for Applied Scientific Research (TNO), Amsterdam, NL. His current research interests include the optimization, planning, and control of energy systems, energy markets, and machine learning applied to energy systems. 
\end{IEEEbiographynophoto}

\begin{IEEEbiographynophoto}{Nataly Ba\~{n}ol Arias} received the B.Sc. degree in Production Engineering from the Universidad Tecnológica de Pereira, Colombia in 2012, and the M.Sc. and Ph.D. degree in Electrical Engineering from the São Paulo State University (UNESP), Ilha Solteira, Brazil, in 2015 and 2019, respectively. Currently, she is a researcher at the University of Twente, The Netherlands. Her current research interests include the development of methodologies for the optimization, planning, and control of modern distribution systems including electric vehicles and renewable energy sources, energy management systems, and flexibility markets.
 
\end{IEEEbiographynophoto}

\begin{IEEEbiographynophoto}                    {Pedro P. Vergara} was born in Barranquilla, Colombia in 1990. He received the B.Sc. degree (with honors) in electronic engineering from the Universidad Industrial de Santander, Bucaramanga, Colombia, in 2012, and the M.Sc. degree in electrical engineering from the University of Campinas, UNICAMP, Campinas, Brazil, in 2015. In 2019, he received his Ph.D. degree from the University of Campinas, UNICAMP, Brazil, and the University of Southern Denmark, SDU, Denmark, funded by the Sao Paulo Research Foundation (FAPESP).  In 2019, he joined the Eindhoven University of Technology, TU/e, in The Netherlands as a Postdoctoral Researcher. In 2020, he was appointed as Assistant Professor at the Intelligent Electrical Power Grids (IEPG) group at Delft University of Technology, also in The Netherlands. His main research interests include the development of methodologies for control, planning, and operation of electrical distribution systems with high penetration of low-carbon energy resources (e.g, electric vehicles, PV systems, electric heat pumps) using optimization and machine learning approaches. Dr. Vergara has received the Best Presentation Award at the Summer Optimization School in 2018 organized by the Technical University of Denmark (DTU) and the Best Paper Award at the 3rd IEEE International Conference on Smart Energy Systems and Technologies (SEST), in Turkey, in 2020.
\end{IEEEbiographynophoto}

\begin{IEEEbiographynophoto}{Maria Vlasiou} is a Professor at the University of Twente, The Netherlands, an Associate Professor at the Eindhoven University of Technology (TU/e), and Research Fellow of the European research institute EURANDOM. She received her B.Sc. (2002, Hons.) and Ph.D. (2006) from the Aristotle University of Thessaloniki and TU/e, respectively. In 2006, she moved to the H. Milton Stewart School of Industrial and Systems Engineering, at the Georgia Institute of Technology, where she first worked as a Research Engineer and later as a Postdoctoral Fellow. Her research interests centre on stochastic processes and stochastic operations research. Her research focuses on the performance of stochastic processing networks with layered architectures and on perturbation analysis for heavy-tailed risk models. Other interests include Lévy processes, large deviations for non-monotone stochastic recursions, and proportional fairness in heavy traffic for bandwidth-sharing networks. She has supervised six PhD theses on these topics. Prof. Vlasiou has been invited to more than 20 foreign universities for collaboration and seminars. She has been associate editor in four journals and has refereed for about 45 international journals, conferences, and national science foundations. Prof. Vlasiou’s research so far has been funded by grants from more than 10 science foundations, universities, societies, and organisations. She is the co-author of more than 50 refereed papers, the co-recipient of the best paper award in ICORES 2013, the Marcel Neuts student paper award in MAM8, a prize at the 8th conference in Actuarial Science, and the recent winner of the INFORMS UPS G. Smith award.
 
\end{IEEEbiographynophoto}

\begin{IEEEbiographynophoto}{Gerwin Hoogsteen} received the PhD degree from the University of Twente in 2017 with his thesis “A Cyber-Physical Systems Perspective on Decentralized Energy Management”. He is currently employed as permanent researcher in the field of smart grids within the Computer Architecture for Embedded Systems chair, with a focus on applying theoretical research in field-tests. His research interest is in energy management for smart grids, and in particular where it concerns multi-disciplinary research and cyber-physical systems. Current research directions include the use of machine learning and artificial intelligence in smart grids, distributed coordination, and cyber-security of smart grids. Hoogsteen is the founder and maintainer of the DEMKit and ALPG software.
 
\end{IEEEbiographynophoto}

\begin{IEEEbiographynophoto}{Johann Hurink} received the Ph.D. degree from University of Osnabrück (Germany) in 1992 for a thesis on a scheduling problem occurring in the area of public transport. Since 2009 he is a full professor at the University of Twente and since 2020 also the Director of the 4TU Applied Mathematics Institute (AMI) in The Netherlands. He has published more than 190 refereed papers in international journals and conferences and has been involved in many European and national research projects. Current research mainly focuses on optimization and control problems for energy management and smart grids.
\end{IEEEbiographynophoto}

\end{document}